\documentclass[journal]{new-aiaa} 
\usepackage[utf8]{inputenc}

\usepackage[version=4]{mhchem}
\usepackage{siunitx}
\usepackage{longtable,tabularx}
\usepackage{subcaption}
\usepackage{comment}
\usepackage{footnpag}			      	
\usepackage{multirow}
\usepackage{booktabs}
\usepackage{hyperref}

\usepackage{makecell}
\usepackage{adjustbox}
\usepackage[numbers]{natbib}
\usepackage[textwidth=2cm,textsize=scriptsize]{todonotes}
\usepackage{nomencl}
\makenomenclature

\begin{document}
\title{Simulating the Evolution of Lethal Non-Trackable Population and its Effect on LEO Sustainability}
\author{ Daniel Jang \footnote{\textit{Corresponding Author}. Ph.D. Candidate, Department of Aeronautics and Astronautics, Massachusetts Institute of Technology, MA 02139, USA. email: djang@mit.edu}}
\author{Richard Linares \footnote{Rockwell International Career Development Professor, Associate Professor of Aeronautics and Astronautics, Department of Aeronautics and Astronautics, Massachusetts Institute of Technology, Cambridge, MA 02139, USA. email: linaresr@mit.edu}}
\affil{Massachusetts Institute of Technology, Cambridge, MA 02139}

\maketitle
 
\begin{abstract}
The vast majority of the orbital population today is unobservable and untracked because of their small size.  These lethal non-trackable objects will only become more numerous as more payloads and debris are launched into orbit and increase the collision rate.  In this paper, the long-term effect of collisions is simulated with an efficient Monte-Carlo method to simulate the future LEO environment including lethal non-trackable objects, which is typically ignored due to the large computational resources required.  The results show that simulations that do not incorporate lethal non-trackable debris would be omitting a large number of debilitating collisions with active payloads and catastrophic collisions to a smaller effect.  This shows the importance of simulating small debris in the long-term evolution of the orbital population, which is often omitted in the literature.  This increased debris population and consequentially the risk to orbital payloads diminishes as smaller lethal non-trackable objects are considered.  An efficient and validated model is used to simulate these numerous small objects. Several future cases such as launches of registered megaconstellations, improved post-mission disposal rates and no-future launches are explored to understand the effect of the inclusion or exclusion of lethal non-trackable objects.
\end{abstract}

\nomenclature[A]{}{}
\nomenclature[A]{AMR}{Area to Mass Ratio}
\nomenclature[A]{ESA}{European Space Agency}
\nomenclature[A]{FCC}{Federal Communications Commission}
\nomenclature[A]{GEO}{Geosynchronous Orbit}
\nomenclature[A]{ITU}{International Telecommunication Union}
\nomenclature[A]{LEO}{Low Earth Orbit (LEO)}
\nomenclature[A]{LNT}{Lethal Non-Trackable}
\nomenclature[A]{MASTER}{ESA's Meteoroid And Space debris Terrestrial Environment Reference}
\nomenclature[A]{MC}{Monte Carlo}
\nomenclature[A]{MOCAT}{MIT Orbital Capacity Analysis Tool}
\nomenclature[A]{SSN}{Space Surveillance Network}
\nomenclature[A]{CAM}{Collision Avoidance Maneuver}
\nomenclature[A]{PMD}{Post Mission Disposal}

\nomenclature[B]{$A$}{Area}
\nomenclature[B]{$a$}{Semi-major axis}
\nomenclature[B]{$f$}{True anomaly}
\nomenclature[B]{$e$}{Orbit eccentricity}
\nomenclature[B]{$i$}{Inclination}
\nomenclature[B]{$J_2$}{Earth's oblateness}
\nomenclature[B]{$L_C$}{Characteristic length}
\nomenclature[B]{$m$}{Mass}
\nomenclature[B]{$P_C$}{Probability of collision}
\nomenclature[B]{$t$}{Time}
\nomenclature[B]{$\Omega$}{Right Ascension of the Ascending Node (RAAN)}
\nomenclature[B]{$\omega$}{Argument of perigee}
\nomenclature[B]{$M$}{Mean anomaly}
\nomenclature[B]{$R_E$}{Earth radius}
\nomenclature[B]{$\rho$}{Atmospheric density}
\nomenclature[B]{$r$}{Radius of object}
\nomenclature[B]{$v_c$}{Collision velocity}
\nomenclature[B]{$s_i$}{Spatial density of objects of type $i$}
\nomenclature[B]{$dU$}{Inverse of the volume of cube}
\nomenclature[B]{$\alpha$}{Collision avoidance failure rate between active and non-active objects}
\nomenclature[B]{$\alpha_a$}{Collision avoidance failure rate between active objects}
\nomenclature[B]{$P_{PMD}$}{Probability of succesful PMD}
\nomenclature[B]{$\Delta t$}{Simulation time step}
\printnomenclature

\section{Introduction} 
Our knowledge of the orbital objects in LEO is best understood for the objects that are measured regularly. 
This usually means that the objects are tracked and have orbits attributed to them with some fidelity such that sensors are able to reacquire and constantly update the state. Characterization of such objects can be done using various phenomenologies such as multispectral imaging, radar imaging, and even passive RF tracking for objects that are emitting in the RF spectrum.  Ever since the launch of Sputnik in 1957, the US government has led the effort to track orbital objects. The Millstone Hill Radar under development for the planned Ballistic Missile Early Warning System by Lincoln Laboratory in Massachusetts was the first U.S. radar to detect and track Sputnik 1.  

Searching and tracking objects in orbit has historically been done using terrestrial active and passive sensors such as radars and telescopes.  Technological progress since then has expanded measurement methods to include passive radio frequency and remote sensing from space.  Due to the resources needed to keep custody of these objects, government entities such as the US Space Surveillance Network (SSN) and the 18th Space Defense Squadron, European Space Operations Centre in ESA have traditionally tracked the objects.  In the past couple of decades, a thriving commercial SSA sector has been introduced. The global space situational awareness market is projected to grow from \$1.3 billion in 2022 to nearly \$2 billion in 2028 with US companies \cite{ssaMarket}.  

Although tracking intact payloads and rocket bodies in LEO has largely become routine, tracking small debris remains a challenge.  Not only is detection itself difficult for small objects due to sensitivity limitations of the SSA sensors, the number of objects grows exponentially as the limit of detectable size gets smaller, which leads to difficulties with identifying the detected object to create a track and an orbit, an \textit{association} problem. The US Space Surveillance Network is understood to be able to track objects larger than 10 cm.  

Tracking and maintaining the custody of small debris to keep an orbit may be difficult; however, routine measurements of the density of debris have been made over the years.  For example, radar measurement data from the Haystack Ultra-wideband Satellite Imaging Radar (HUSIR) is provided to the NASA Orbital Debris Program Office for measurements of objects down to 3 mm \cite{Stansbery2008, haystack2022}.  The radar operates at a beam-park mode with a fixed elevation and azimuth pointing direction to obtain the number of objects, RCS and doppler measurement of the objects that cross through the field of view.  This provides a fixed detection volume that simplifies calculations of the debris flux, or number of objects detected per unit area, per unit time.  The effects of collision events and ASAT tests have been reported by statistical sampling of the population and the characteristics of the resulting collision debris \cite{ODQN-2007-07, ODQN-2009-04, ODQN-2010-04, ODQN-2011-10, ODQN-2023-10, ODQN-2024-04}.

In situ measurements of the small debris population have also been made in LEO.  Debris as small as several microns in size have been measured by impact craters onto space-exposed materials from satellites and specific missions with sensors to measure on-orbitdebris impacts.  Although specific orbital information for individual debris is difficult to measure, the statistical distribution of objects with respect to their orbital parameters and physical characteristics can be estimated.  The Long Duration Exposure Facility was launched in April 1984 by the Space Shuttle mission STS-41-C at an altitude of about 480 km, where it sampled the LEO environment for more than 2000 days. The LDEF (Long Duration Exposure Facility) mission was in orbit between 1984 and 1990 and showed that there was a large amount of debris in the LEO environment with a population following the power law and showed the large amount of debris that eluded terrestrial sensors \cite{McDonnell1993TheOverview, Durin2022ActiveSatellite}.  The replaced solar arrays from the Hubble Space Telescope (HST) Service Missions SM-1 and SM-3B in 1993 and 2002, respectively, have provided the space community with solar arrays that have been exposed to the space environment at about 600 km for more than seven years \cite{MOUSSI20051243, KEARSLEY20051254, 382250}. Other in situ measurements have been conducted since then.  DEBIE-1 (DEBris In orbit Evaluator) was launched on a small technology satellite PROBA in 2001 \cite{DEBIE1, Drolshagen2006}, and DEBIE-2 was launched in 2008 as part of the European Technology Exposure Facility (EuTEF) and installed on the exterior of the ISS \cite{DEBIE2}. The Space Debris Sensor (SDS) is currently mounted on the exterior of the International Space Station \cite{SDS2017, SDS2019, 7557063}.  Since 2017, the SOLID sensor system onboard a small satellite TechnoSat has been measuring the impact of debris \cite{BAUER2022235}.

Many debris models have been created using these data for the past and future LEO orbital debris environment, including NASA's ORDEM \cite{ORDEM2019vnv,ORDEM2022model} and ESA's MASTER \cite{MASTER8}.  The debris population from these databases for some initial simulation epoch can be used to seed the initial population with randomized orbits.  The difficulty lies in the fact that for smaller object sizes, the number grows rapidly.  ESA estimates that there are more than 1 million fragments between 1 and 10 cm and around 130 million objects between 1 mm and 1 cm in orbit around Earth as of June 2023.  
Any impact with one of these objects threatens to at least impair the functioning of a working spacecraft, or at worst destroy it altogether, creating ever more debris.  

The risk to orbiting spacecraft from the sub-cm size meteoroids and space debris particle has been studied in literature, with demonstrations showing how these small particles may seriously damage or destroy payloads \cite{ADUSHKIN2020591, DROLSHAGEN20081123}.  The computational power needed to simulate an environment with that many objects is difficult, and thus most MC models have limited analysis to $>10$ cm objects, or run specific cases with smaller objects.  Not only does the initial population need to be propagated, with a smaller size threshold for simulations, the debris generated from collisions will also produce objects as small as that designated threshold.  

Statistical sampling methods propagate every object's orbital states with high fidelity propagators to estimate the future space environment at some small time steps, much like a particle filter.  
Several such sampling-based models have been developed by space agencies and private entities as a result of the large-scale development and support required.  Examples include NASA’s Orbital Debris Engineering Model (ORDEM) and LEO to Geosynchronous Orbit Debris model (LEGEND)\cite{LEGEND2004}, European Space Agency’s Orbital Debris Evolutionary Model (ODEM), Chinese Academy of Sciences’ SOLEM (Space Objects Long-term Evolution Model), University of Southampton and United Kingdom Space Agency's Debris Analysis and Monitoring Architecture for the Geosynchronous Environment (DAMAGE)\cite{Lewis2001, Lewis2011}, MEDEE model from Centre National d’Etudes Spatiales \cite{medee}, DELTA model from European Space Agency \cite{DELTA2}, LUCA model from Technische University at  Braunschweig \cite{luca2}, NEODEEM model from Kyushu University and the Japan Aerospace Exploration Agency, IMPACT \cite{Sorge2014IMPACTInvited}, and others \cite{Wang2019AnSOLEM, Drmola2018KesslerModel, Rosengren2019DynamicalOrbits}.  A probabilistic debris environment propagator has been explored in \cite{Giudici2024} where objects are classified as either intact objects or fragments and accounts only for collisions between the two classes.  This simplification allows for computational efficiency and results show a good match for the trackable debris population, though validation work is on-going.

Several contributions to the literature are presented in this paper.  MOCAT-MC, an efficient MC method \cite{jang2024monte}, is leveraged to simulate the high population of LNTs, which has been a limiting factor in space debris research due to the enormous computational resources typically required.  The effect of LNTs for several scenarios is compared, including \textit{No Future Launch} scenario and potential future megaconstellation launch scenario.  The inclusion of LNTs in the evolution of the LEO population is shown to be important, and the omission of such population in any simulation paints a more optimistic result with regard to collisions and population growth in LEO.  The relative increase in catastrophic and noncatastrophic collisions is compared and discussed.  The large increase in noncatastrophic collisions shows that the increase in the LNT population can be seen as a degradation of Post Mission Disposal (PMD) capability.

\section{Modeling the effect of Lethal Non-Trackable Objects}

In this paper, LNT is defined as objects that are sufficiently small to not be reliably detected to be actionable for collision avoidance maneuvers by the satellite operators and the payload.  The LNT population is quickly populated from any simulation that allows small objects to be created from collisions, and can also be initialized from other debris density models such as NASA ORDEM or ESA MASTER.  The effect of this population is modeled in the newly modified MOCAT-MC model to add additional functionality to support LNTs, leveraging its ability to simulate millions of objects with ease.

\subsection{Fragmentation Model}
 
Fragmentation events in MOCAT-MC are simulated with the NASA Standard Breakup Model (SBM) \cite{nasaSTMevolve4, KlinkradBook2006}, which is summarized below.  The history of the technical development of breakup models and alternative methods is extensively outlined in the MASTER user guide \cite{MASTER8}. The SBM is a semiempirical model based on evidence compiled from historical orbital data and terrestrial laboratory experiments.  The model is deterministic and sample-based, and the samples are described by $L_C$ the characteristic length, AMR the area-to-mass ratio, and $\Delta v$ the ejection velocity in a random direction from the parent velocity. 

The SBM specifies that the impact energy per target mass, or specific energy, is defined as
\begin{equation}
    \epsilon = \frac{1}{2}\frac{m_c}{m_t}v_c^2,
\end{equation}
where $m_c$ is the mass of the chaser and $m_t$ is the mass of the target, and impact velocity is $v_c$.  The mass of the target is assumed to be greater than the mass of the chaser.  A collision is considered to be catastrophic where the chaser and the target are completely fragmented when $\Tilde{E}_p^*=40$  J/g. 
The model is initialized by attributing the characteristic length $L_C$ which acts as the smallest size for the objects considered. 
 This is a variable that can be adjusted, though MOCAT-MC's default value is 0.1 m. Note that $L_C$ will be treated as an equivalent diameter $d$. 

\subsection{Collision Avoidance Maneuver Modeling}

In MOCAT-MC, the LNT population is defined as objects with $L_C < 10$ cm.  The model takes into account $\alpha,\alpha_a$ terms that modifies the probability of collision between an active payload and a derelict object, and between two active payloads, respectively.  These terms account for the efficacy of the collision avoidance maneuver (CAM), since the maneuver can be planned.  Although this is a static value effective for all tracked objects, it is modified for LNTs whose orbital states are either not known or are tracked with larger uncertainties.  

\begin{figure}[htb]
    \centering
    \includegraphics[width=0.5\textwidth]{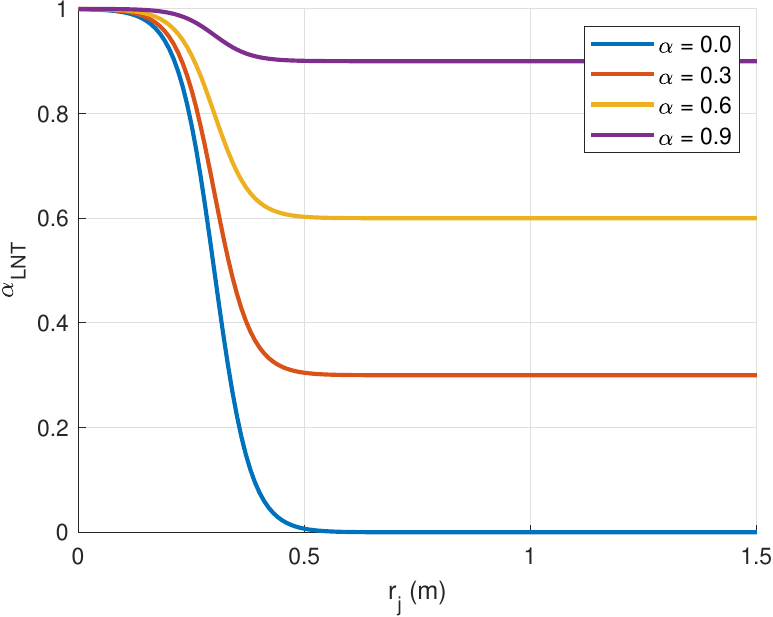}
    \caption{The collision avoidance efficacy term $\alpha_{LNT}$ for LNT objects for a range of baseline efficacy $\alpha$}
    \label{fig:LNTalphas}
\end{figure}

The efficacy of CAMs against LNT objects is assumed to follow a logistic curve.  With this modifier, the $\alpha$ term is modified such that $\alpha$ effectively becomes 1 as the radius of the inactive object nears 0, effectively eliminating any reduction of the probability of collision.  The modified $\alpha$ for LNT objects is as follows: 

\begin{align}\label{eq:LNTalpha}
    \alpha_{LNT} = \left(1- \frac{1}{1+ \exp\left(-25(r_j-0.3)\right)}\right)(1-\alpha)+\alpha,
\end{align}

where $r_j$ is the radius of the nonactive object and $\alpha$ is the original collision avoidance term.  A range of $\alpha_{LNT}$ is shown in Fig. \ref{fig:LNTalphas} using Eq. \ref{eq:LNTalpha} as the modifier. The $\alpha$ term is maintained for large objects and moves towards 1 as the radius of the object gets smaller and evades detection. 

Note that all payloads are assumed to be well-tracked; therefore, the $\alpha_a$ term is not modified.  At the time of writing, no two maneuverable and controlled active satellites had collided, effectively rendering $\alpha_a = 0$.  With the continued increase in the payload population in LEO, this term can be modified to reflect a more realistic collision avoidance metric.  The description and discussion of the other aspects of the MOCAT-MC model can be found in \cite{jang2024monte}. 
\section{Validation with ADEPT dataset}

Aerospace Corpopration's evolutionary model called Aerospace Debris Environment Projection Tool (ADEPT) \cite{ADEPT2020} is used as a benchmark for the validation of MOCAT-MC's LNT performance.  ADEPT is capable of propagating objects between LEO and GEO, and uses the Orbit Crossing method for collision calculation and an internal fragmentation model called IMPACT \cite{Sorge2014IMPACTInvited}.  ADEPT is also able to group homogeneous objects in similar orbits into one object with some weighting factor to represent multiple objects - sometimes numbering in the thousands.  This method allows ADEPT to propagate one representative object instead of having to do so for all the represented objects.  An initial population is given (\textit{popZero}), and the Future Launch Model (\textit{FLM}) can be specified with a mix of Continuously Replenishing Constellations (\textit{CRC}), non-replenishing constellations (\textit{Non-CRCs}), and Future Constellation Models (\textit{FCM}).  All objects simulated in ADEPT are defined by 16 attributes, as described in Table \ref{tab:ADEPTdefinitions}.  

\begin{table}[htp!]
\centering
\caption{Definition of parameters for each orbital object in ADEPT}
\begin{tabular}{@{}lccccc@{}}
\toprule
Column & 1  & 2  & 3-8  & 9  & 10  \\ \midrule
Description & ID &  Start epoch & $\bar{a},\bar{e},\bar{i},\bar{\Omega},\bar{\omega},\bar{M}$  & End epoch  & Object type \\  \bottomrule  \toprule
Column & 11    & 12-13 & 14 & 15 &  16 \\ \midrule
Description & Disposal flag  & Stationkeeping flag & Area and mass & Diameter & Weighting \\
\bottomrule
\end{tabular}
\label{tab:ADEPTdefinitions}
\end{table}

``Historical'' PMD success rates were used, where PMD = 90\% for CRC LEO, GEO, MEO, and LLC satellites and PMD = 70\% for nonCRC LEO satellites.  A successful PMD results in the object being removed from the environment; whereas a failed PMD would leave it in the simulation.  Operational satellites were assumed to avoid collisions with > 10 cm fragments with 100\% efficacy.  

A dataset of input and output populations and statistics was obtained for the use of MOCAT-MC validation.  The simulation for the dataset had a starting epoch of Dec 1, 2022, and the outputs for 100 runs were provided.  The initial population given had 399,399 objects, and the operational payload count was steady at 18,006 objects.  In the given ADEPT dataset, many objects could be represented by one object where the weighting is $>1$.  For example, after a fragmentation event, ADEPT's collision model IMPACT creates a few representative object sizes with multiple weighting factors that span down to 2 cm objects.  This weighting factor can also be a decimal.  In the conversion to the MOCAT simulation, all of these objects are independently represented with a randomized mean anomaly.  For the cases of noninteger weighting factors, a random value is used between the floor and the ceiling of that factor.  The SBM for MOCAT was run with $L_C=2$ cm. 

\begin{figure}[!ht]
    \centering
	\subcaptionbox{Initial population}[.45\textwidth]{\includegraphics[width=0.9\linewidth]{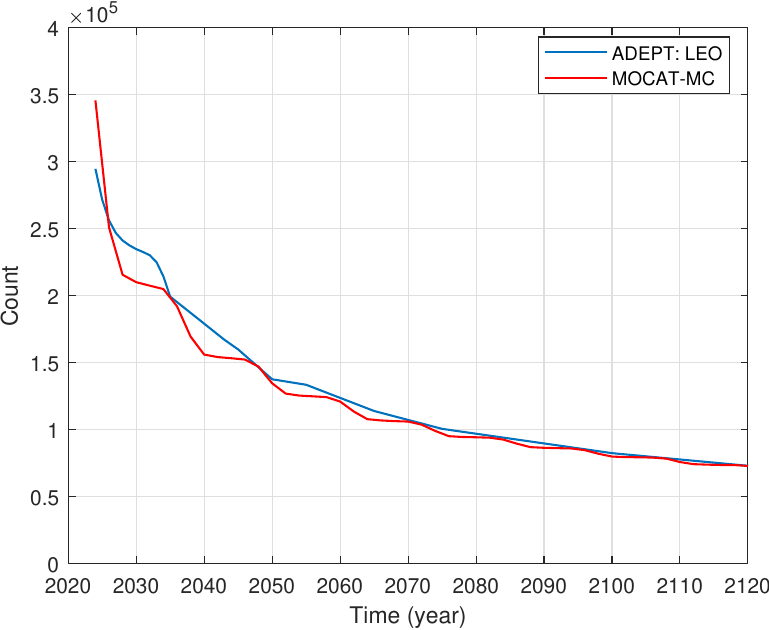}}    \subcaptionbox{Payload population}[.45\textwidth]{\includegraphics[width=0.9\linewidth]{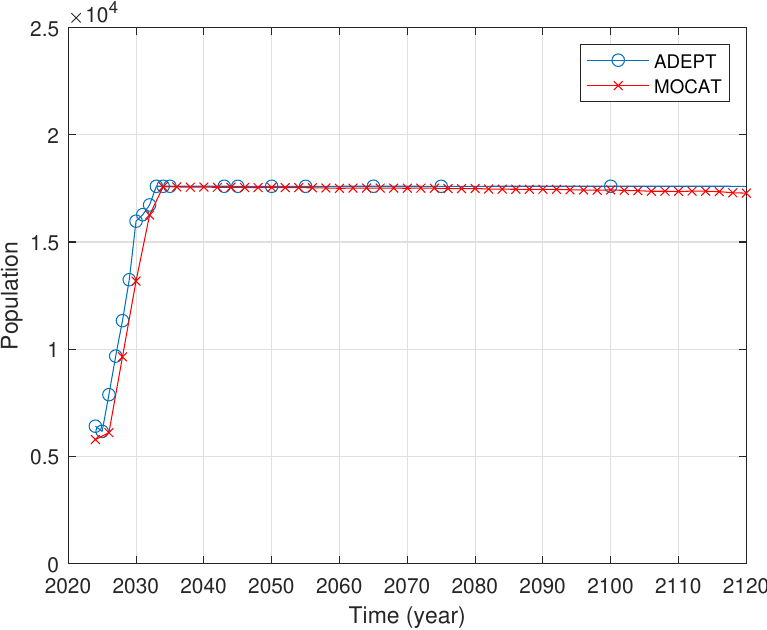}}	
	\caption{Population comparison for ADEPT vs MOCAT-MC}
	\label{fig:ADEPTvnvPopzero}
\end{figure}

The comparison of the initial population reduction between the two models for this scenario is shown in Fig. \ref{fig:ADEPTvnvPopzero}.  For MOCAT, the mean of 10 runs is shown for all the results.  This shows good agreement between the two models, which validates the propagator and the drag model used in MOCAT-MC.  The waviness of the population is due to the time-varying atmospheric density, and the phasing difference between the two models.  The active payload count is also shown in the figure, which shows good agreement for the launch and PMD model.  The launch traffic varied between 5500 and 8000 objects per year after 2030, and included payloads (controlled and uncontrolled), rocket bodies, and debris.

The comparison of the initial population reduction between the two models for this scenario is shown in Fig. \ref{fig:ADEPTvnvPopzero}.  For MOCAT, the mean of 10 runs is shown for all the results.  This shows good agreement between the two models, which validates the propagator and the drag model used in MOCAT-MC.  The waviness of the population is due to the time-varying atmospheric density and the phasing difference between the two models.  The active payload count is also shown in the figure, which shows good agreement for the launch and PMD model.  Launch traffic varied between 5500 and 8000 objects per year after 2030, and included payloads (controlled and uncontrolled), rocket bodies, and debris.    
This validation exercise shows that MOCAT and ADEPT agree well in the propagation module, atmospheric model, and launch model even for LNT objects down to the 2 cm objects.  


\section{Results}
\subsection{No Future Launch scenarios}

Even if objects smaller than 10 cm do not interact with any other objects, the number of objects in that region produced from collisions would show a higher number compared to a scenario that allows only for objects greater than 10 cm.  Of note is the number of collisions that grow with the introduction of LNTs.  In order to investigate the growth in the collision rate and population when LNTs are considered, the \textit{No Future Launch} case is run where no more launches are allowed after the scenario epoch of Jan 1 2023, while allowing for LNT debris to exist in the simulation.  The minimum characteristic length of $L_C = \{1,3,5,10\}$ cm is used and compared here.  

\begin{figure}[!ht]
    \centering
	\subcaptionbox{Initial and final population per altitude}[.45\textwidth]{\includegraphics[width=0.9\linewidth]{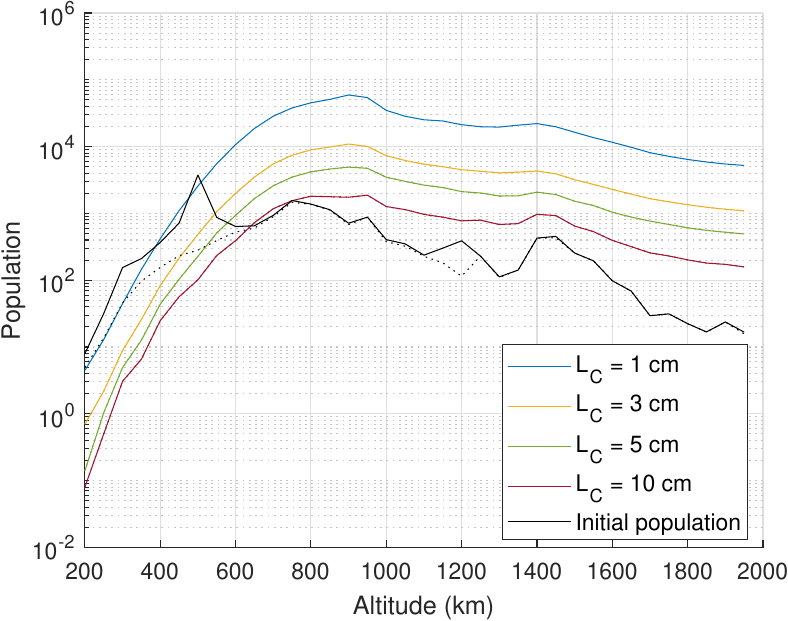}}    \subcaptionbox{Total population}[.45\textwidth]{\includegraphics[width=0.9\linewidth]{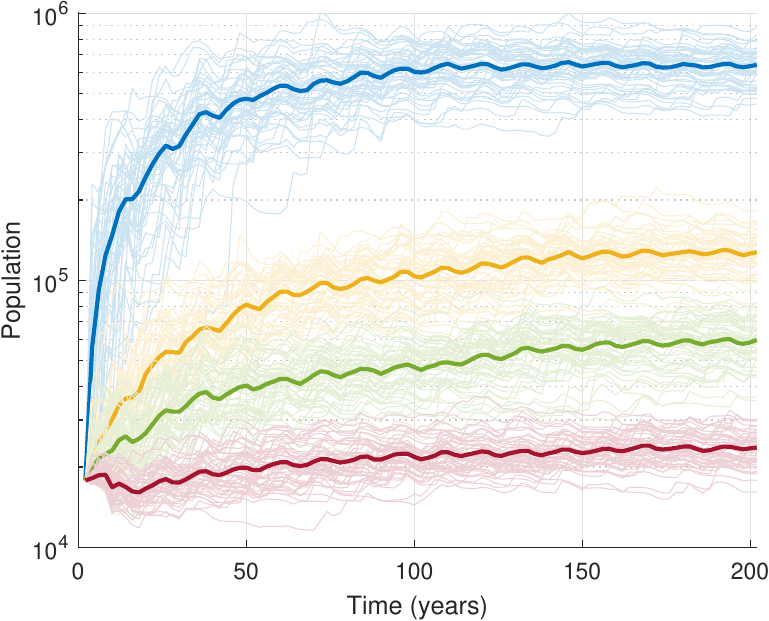}}	
	\caption{Total population count with no future launches over a 200-year span for various minimum LNT sizes ($L_C$). The altitude bins are 50 km.}
	\label{fig:NFLLNT_pop}
\end{figure}

Figure \ref{fig:NFLLNT_pop}(a) shows the altitude distribution of the final total population for the 200-year simulations with various $L_C$ along with the initial population.  The altitude bins are 50 km.  The total population as a function of time is shown in Fig. \ref{fig:NFLLNT_pop}(b).  Consideration of 1 cm sized objects quickly rise to 100's of thousands of objects even in this \textit{No Future Launch} case, and a clear growth of population for all altitudes is shown with reduced $L_C$.  

\begin{figure}[!htb]
    \centering
	\subcaptionbox{Initial and final population size distribution}[.45\textwidth]{\includegraphics[width=0.9\linewidth]{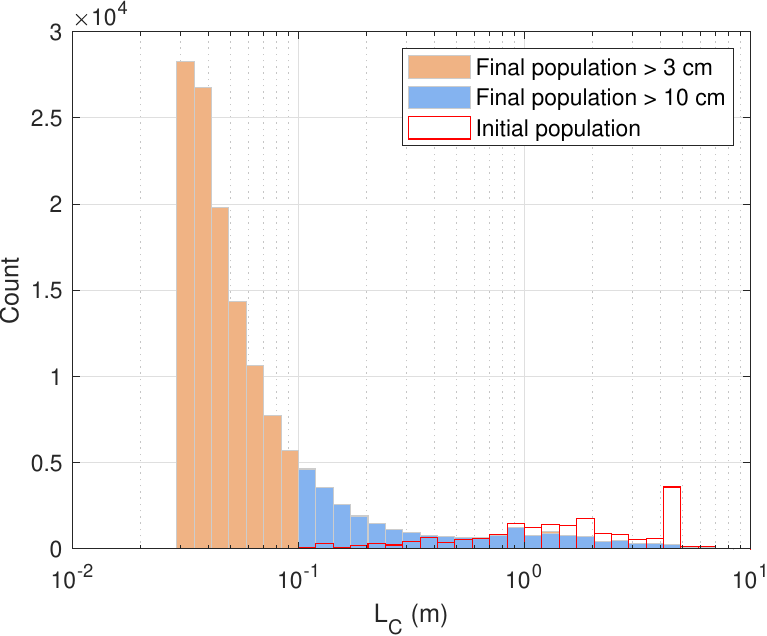}}    \subcaptionbox{Cumulative collisions}[.45\textwidth]{\includegraphics[width=0.9\linewidth]{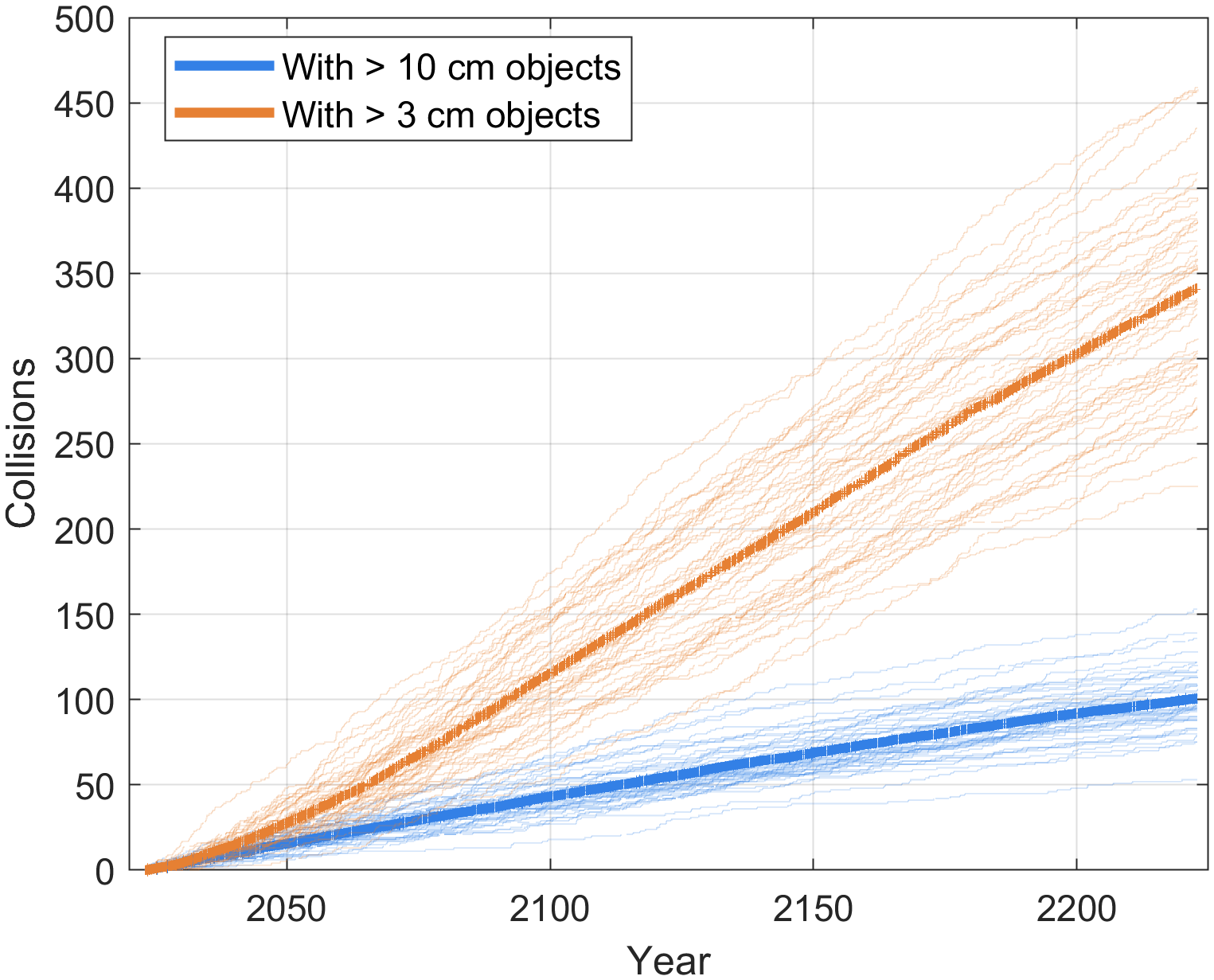}}	
	\caption{No Future Launches scenarios for $L_C$ = 3 cm and 10 cm}
    \label{fig:ESA_NFL_LNT}
\end{figure}

Figure \ref{fig:ESA_NFL_LNT}(a) compares the distribution of the final population at the end of each of the scenario for the $>3$ cm and $>10$ cm cases by size of the object, along with the initial population distribution.  The dotted line denotes the debris population while the solid line denotes the total population.  The red outline also shows the initial population distribution.  Since both are for the \textit{No Future Launch} cases, the initial population's larger objects slowly deorbit or is removed from the simulation due to collision attrition.  The small-object population grows quickly over the 200-year simulation as the debris generated from collisions, especially below the 10 cm region.  It is notable that some bin sizes have fewer population in the $>3$ cm case compared to the $>10$ cm case, especially closer to 10 cm, as these objects also undergo fragmentation events from smaller objects for the $>3$ cm scenarios. In the $>10$ cm scenario, these objects do not collide with other smaller objects simply because they do not exist in the scenario.  This example shows that lowering the characteristic length limit for a simulation replaces the population of trackable objects with LNTs.  Thus, limiting the simulations to only trackable objects gives an incomplete understanding of the LEO environment.  It is also shown here that due to the interaction between the LNTs and smaller-sized objects, a simulation run with LNTs that has been cropped to show the non-LNT objects is not equivalent to a simulation run without considering LNTs.  
Figure \ref{fig:ESA_NFL_LNT}(b) shows the difference in the cumulative number of collisions between two independent sets of scenarios: one allowing objects to exist above 10 cm and another for objects above 3 cm.  Each of the thin line represents one of the 50 MC runs for that scenario, while the thick line is the median value of those runs.  It can be seen that the number of collisions continues to grow even without any new launches into the LEO environment.  For the $> 3$ cm case, the number of collisions grows quicker.  Since collisions result in more collisions, the slopes of these lines have an exponential growth factor for the population.  

Although it is not surprising that there are more collisions with a population that has more objects, an analysis of what types of object are colliding is needed and at what frequency.  For example, with the addition of the 3-10 cm debris objects, the $>$ 3 cm scenario will have many more LNT-on-LNT collisions, which does not directly affect the intact objects.  The probability of collision is affected by the combined cross-sectional area of the two objects, and thus a given pair of LNT objects will not collide as frequently as it would against another larger object.

\subsection{Megaconstellation launches} \label{sec:megaLNTs}

\begin{figure}[!h]
    \centering
	\subcaptionbox{Total operational numbers for megaconstellations}[.45\textwidth]{\includegraphics[width=0.9\linewidth]{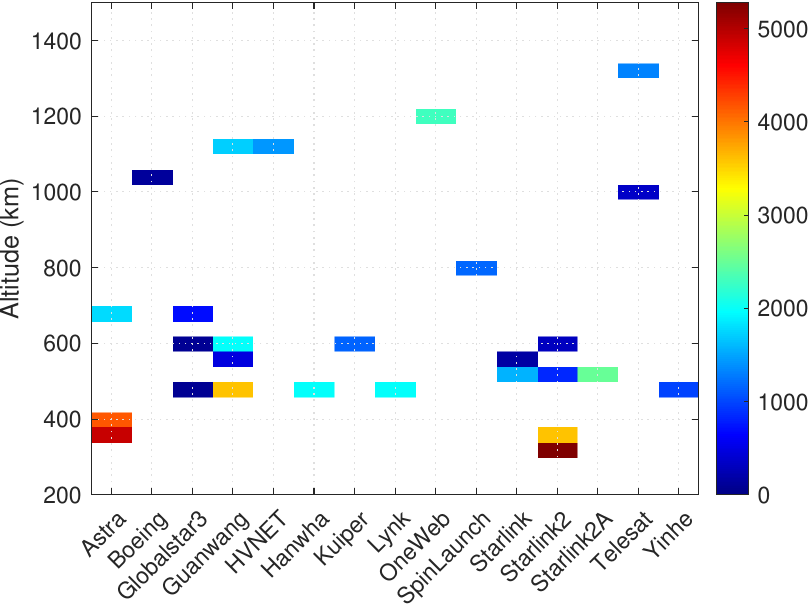}}
	\subcaptionbox{Megaconstellation altitudes}[.45\textwidth]{\includegraphics[width=0.9\linewidth]{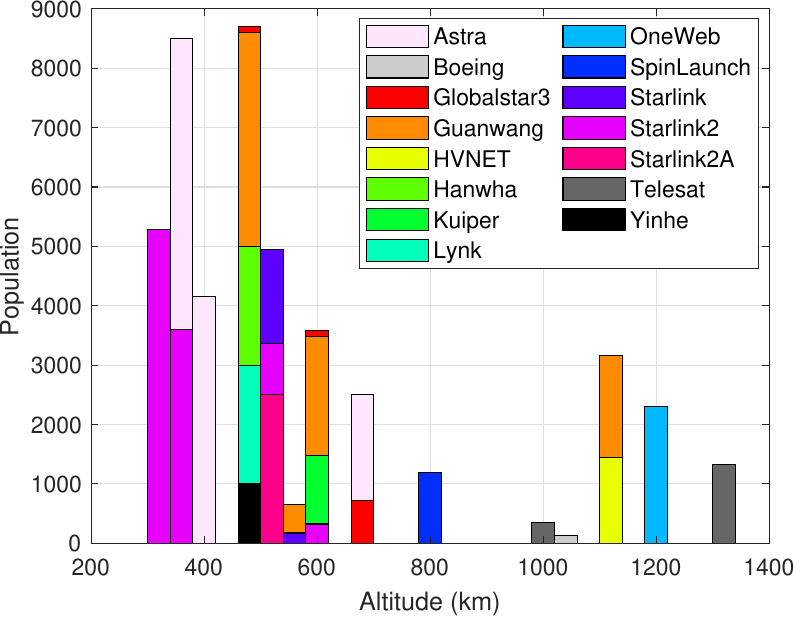}}  
	\caption{Megaconstellation launch altitudes}
	\label{fig:megaTotLaunchAlt}
\end{figure}

The FCC and ITU filed megaconstellation altitudes and sizes are shown in Figure \ref{fig:megaTotLaunchAlt}.  
Though it is impossible to perfectly forecast the LEO traffic and launches in the future, best estimate of the future megaconstellations can be compiled from a few sources, particularly governmental and regulatory bodies such as the FCC and ITU as well as press releases.  As of writing, there are more than 50 megaconstellations -- defined as constellations comprising more than 1000 satellites -- that have been credibly proposed \cite{mcdowell2023jonathan,henry2019spacex,diaz2023data,jang2024monte}.  These are summarized in Fig. \ref{fig:megaTotLaunchAlt} for a total of more than 80,000 satellites in operation.  

The scenarios with a subset of future megaconstellation launches are explored in this section for the varying levels of $L_C$. 
PMD success rates were derived from the historical rates given by ESA, with 40\% for active (non-constellation) payloads, 55\% for rocket bodies, and 90\% for constellation satellites.  Assuming that collision between two active satellites are unlikely, collision avoidance efficacy is chosen to be $\alpha = 0.01$ for actively controlled satellites, while all other cases with two active satellites were set at $\alpha_a = \alpha_{inter} = \alpha_{intra} = 0$.  

\begin{figure}[!ht]
    \centering
        \subcaptionbox{Total population evolution for first 100 years}[.45\textwidth]{\includegraphics[width=0.9\linewidth]{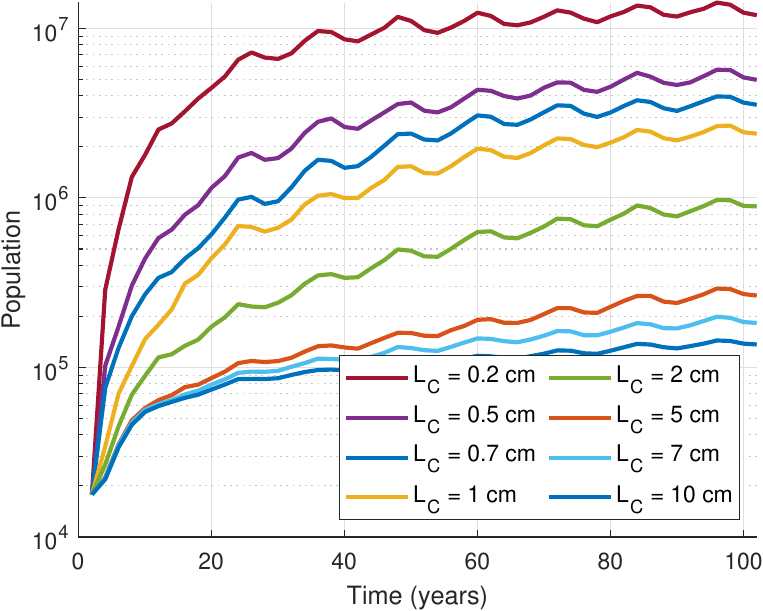}}
	\subcaptionbox{Initial and final population per altitude.  Dotted lines represent all uncontrolled objects.}[.45\textwidth]{\includegraphics[width=0.9\linewidth]{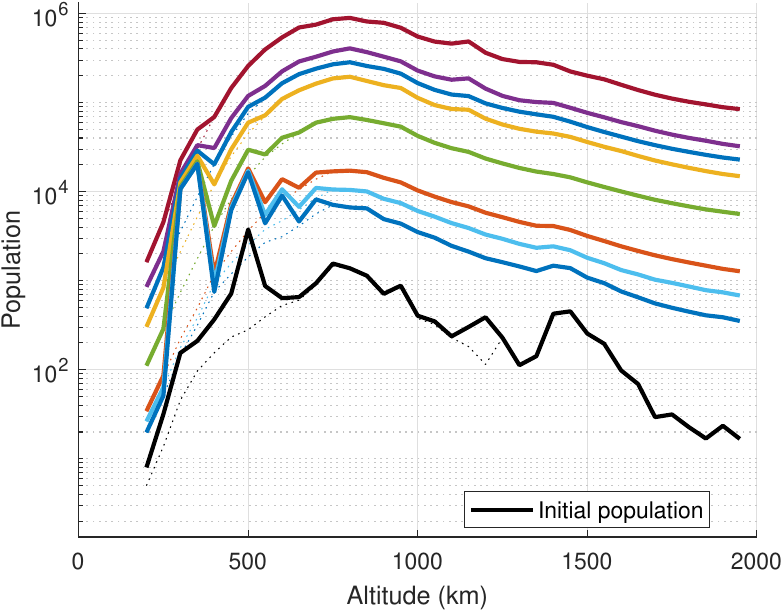}}  
	
    \caption{Total population count with megaconstellations launches limited to $<700$ km over a 200-year span for simulations with $L_C$ between 2 mm and 10 cm}
	\label{fig:mega700LNT_pop}
\end{figure}

\begin{figure}[!ht]
    \centering
    \subcaptionbox{$L_C$ = 2 mm}[.45\textwidth]{\includegraphics[width=0.9\linewidth]{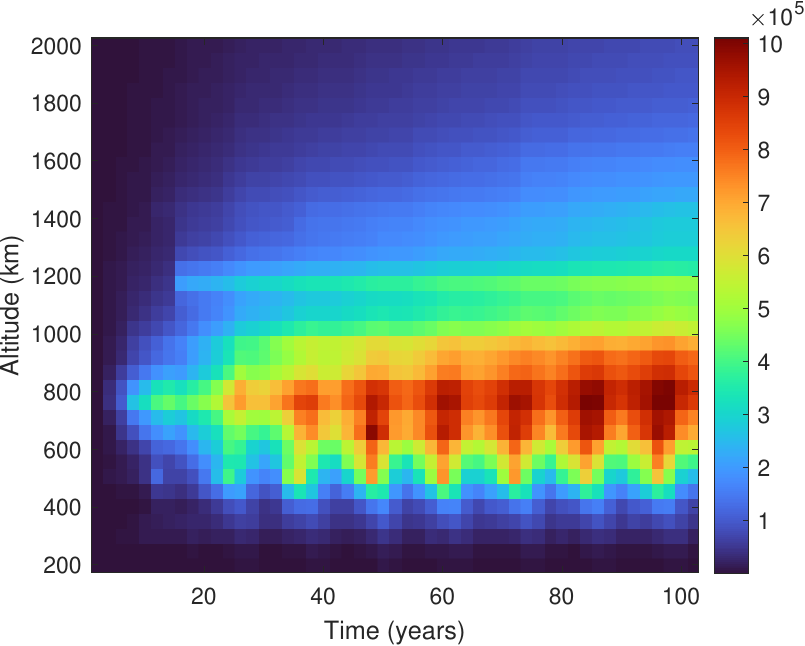}}
	\subcaptionbox{$L_C$ = 7 cm}[.45\textwidth]{\includegraphics[width=0.9\linewidth]{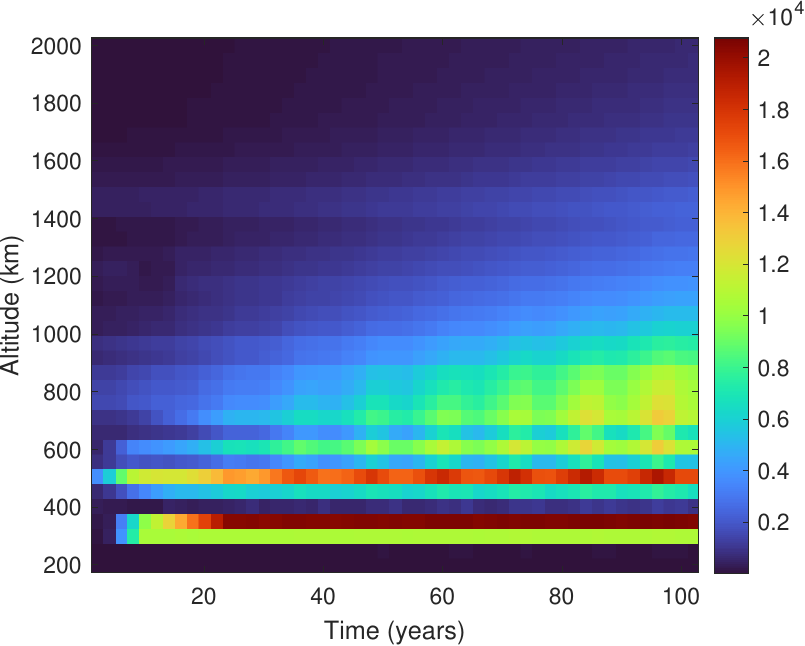}}	
	\caption{Population per altitude with megaconstellations launches limited to $<$ 700 km over a 200-year span for various minimum debris characteristic length $L_C$}
	\label{fig:mega700LNT_imagescPop}
\end{figure}

The launch scenario considers all $<700$ km constellations that total 65,408 operational constellation satellites.  The launch traffic is run with $L_C=\{0.2, 0.5, 0.7, 1, 3,10\}$ cm.  The initial epoch is Jan 1 2023 with a scenario duration of 200 years.  For each $L_C$ value, 20 MC simulations are run.  In this scenario, the number of objects grows as the population in the simulation consists of smaller objects.  The mean values of the MC simulations for each $L_C$ scenario are shown in Fig. \ref{fig:mega700LNT_pop}.  Figure \ref{fig:mega700LNT_imagescPop} shows the population per altitude for $L_C = 7$ cm and 2 mm.  Time is binned at 2 years and altitude is binned at 50 km.  The $L_C = 7$ cm case clearly shows each of the megaconstellation launches, as the vast majority of the objects are constellation objects.  Although no launches are made to $>700$ km altitudes in this scenario, higher altitudes also show an increase in population.  With $L_C=2$ mm, the smaller debris objects are shown whose population is much more numerous compared to the consetllation objects.  The increase in the higher altitudes is shown more clearly and shows up earlier than in the case where smaller objects are omitted.  The atmospheric sink effect is also clearly shown with an orders of magnitude difference in the debris population between high and low solar activity periods for lower altitudes.

\begin{figure}[!ht]
    \centering
	\subcaptionbox{All collisions}[.45\textwidth]{\includegraphics[width=0.9\linewidth]{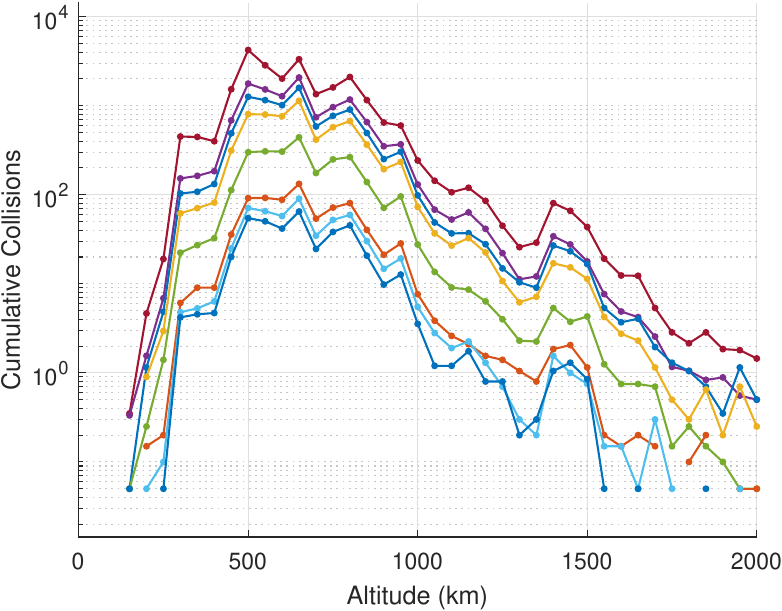}}  
	\subcaptionbox{Catastrophic collisions}[.45\textwidth]{\includegraphics[width=0.9\linewidth]{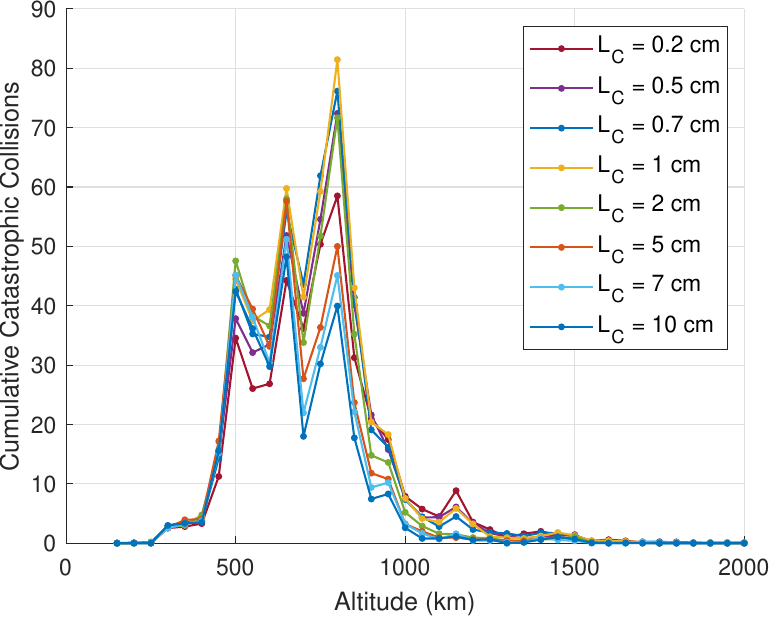}}
    \caption{Cumulative number of collisions with megaconstellations launches limited to $<700$ km over a 200-year span for a range of $L_C$}
	\label{fig:mega700LNT_col}
\end{figure}

The catastrophic and total collision statistics for the three cases are shown in Fig. \ref{fig:mega700LNT_col}, where several conclusions can be drawn.  Figure \ref{fig:mega700LNT_col}(a) shows the total number of collisions during the 200 year simulation (median value is shown per $L_C$ scenario), where the increased number of collision is seen as $L_C$ is reduced.  Figure \ref{fig:mega700LNT_col}(b) shows that the number of catastrophic collisions increases much more slowly with reduced $L_C$, and the catastrophic collision rates are similar for most altitudes.  The altitude with the greatest difference is around 800 km, where the cases of $L_C=0.2$ cm and $L_C=10$ cm differ by approximately a factor of 2.  This is also despite the fact that launches are limited to altitudes $<700$ km, with the highest concentration around 350 km and 550 km.

\begin{figure}[!ht]
    \centering
	\subcaptionbox{$L_C = 10$ cm}[.33\textwidth]{\includegraphics[width=0.95\linewidth]{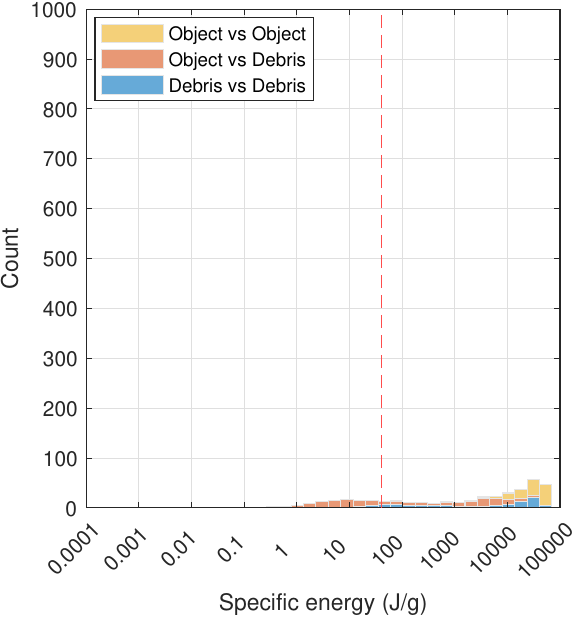}}
	\subcaptionbox{$L_C = 1$ cm}[.33\textwidth]{\includegraphics[width=0.95\linewidth]{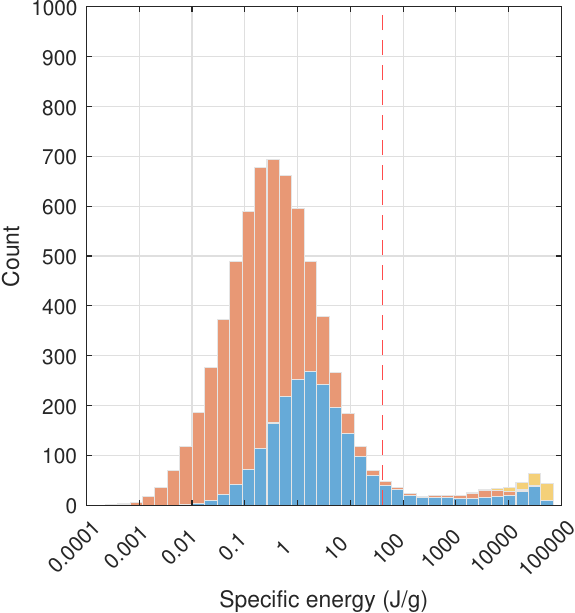}}
    \subcaptionbox{$L_C = 0.2$ cm}[.33\textwidth]{\includegraphics[width=0.95\linewidth]{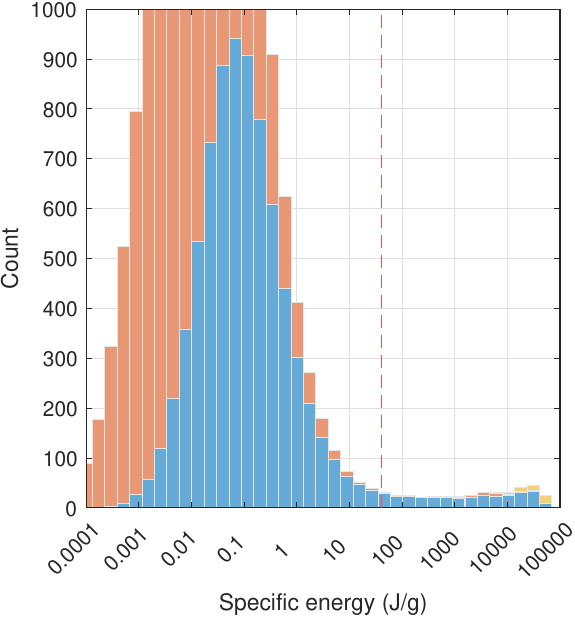}} 
	\caption{Histogram of specific energy of all collisions for megaconstellation launches $<$ 700 km for a range of $L_C$ }
	\label{fig:mega700fi5sStackHist}
\end{figure}

\begin{figure}[!ht]
    \centering
	\subcaptionbox{All collisions}[.33\textwidth]{\includegraphics[width=0.95\linewidth]{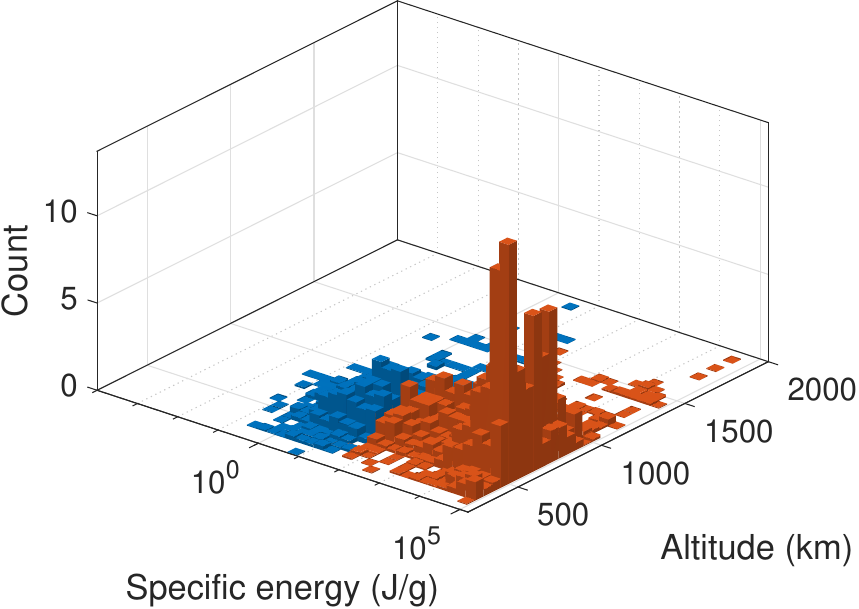}}
	\subcaptionbox{Object on Debris collisions}[.33\textwidth]{\includegraphics[width=0.95\linewidth]{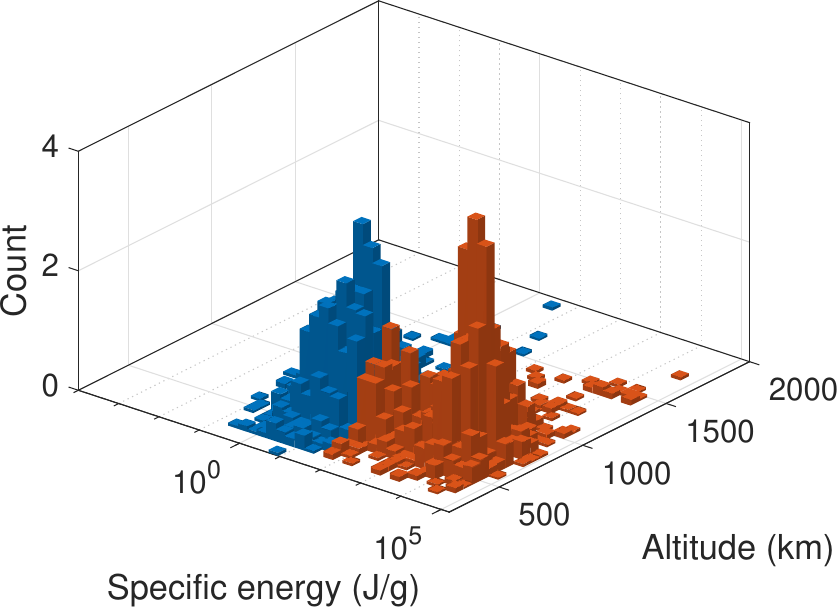}}
 	\subcaptionbox{Object on Debris collisions}[.33\textwidth]{\includegraphics[width=0.95\linewidth]{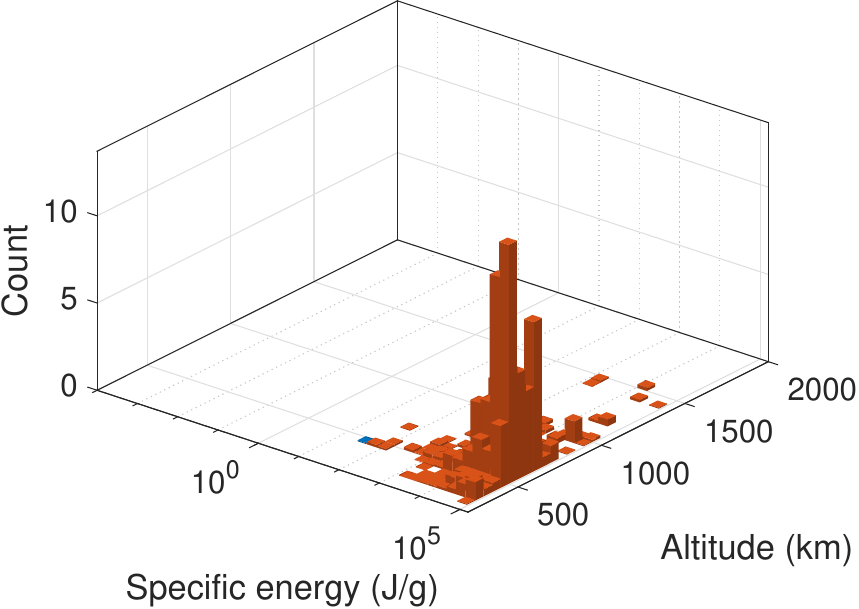}}
  
	\caption{Specific energy of collisions per altitude for a subset of collision population for $L_C = 10$ cm}
	\label{fig:mega700-fi5-10cm}
\end{figure}

\begin{figure}[!ht]
    \centering
	\subcaptionbox{All collisions}[.33\textwidth]{\includegraphics[width=0.95\linewidth]{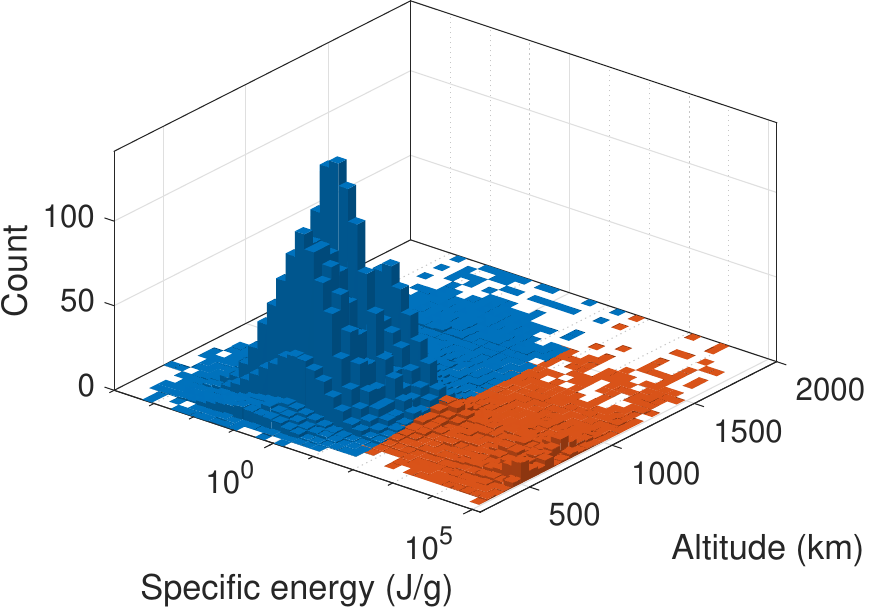}}
	\subcaptionbox{Object on Debris collisions}[.33\textwidth]{\includegraphics[width=0.95\linewidth]{Figures/mega700-fi5-hist3ObjDeb-10cm.pdf}}
 	\subcaptionbox{Object on Debris collisions}[.33\textwidth]{\includegraphics[width=0.95\linewidth]{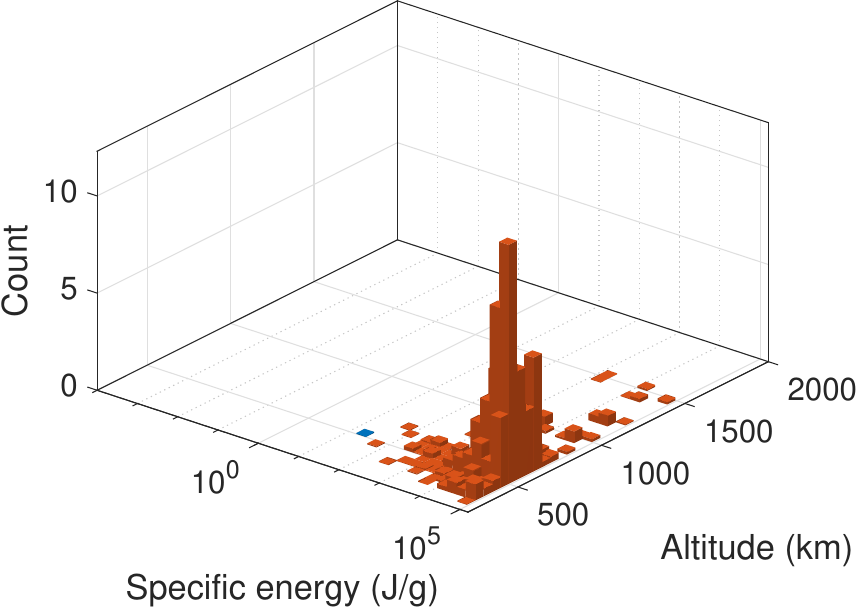}}
  
	\caption{Specific energy of collisions per altitude for a subset of collision population for $L_C = 1$ cm}
	\label{fig:mega700-fi5-01cm}
\end{figure}

\begin{figure}[!ht]
    \centering
	\subcaptionbox{All collisions}[.33\textwidth]{\includegraphics[width=0.95\linewidth]{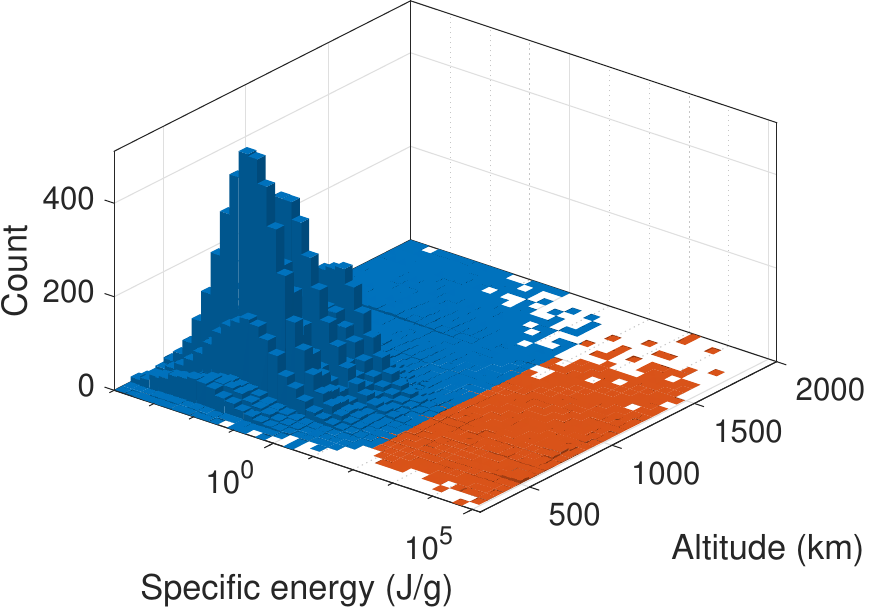}}
	\subcaptionbox{Object on Debris collisions}[.33\textwidth]{\includegraphics[width=0.95\linewidth]{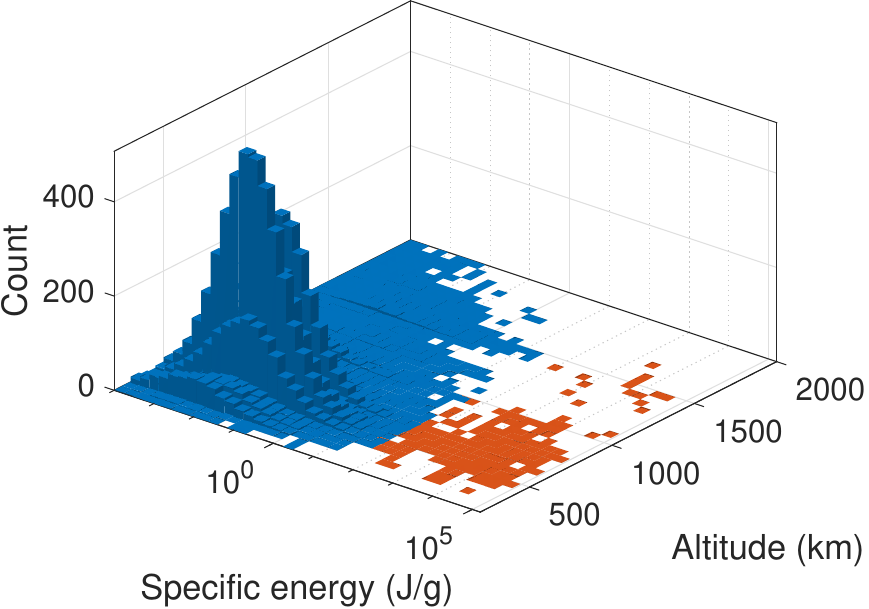}}
 	\subcaptionbox{Object on Debris collisions}[.33\textwidth]{\includegraphics[width=0.95\linewidth]{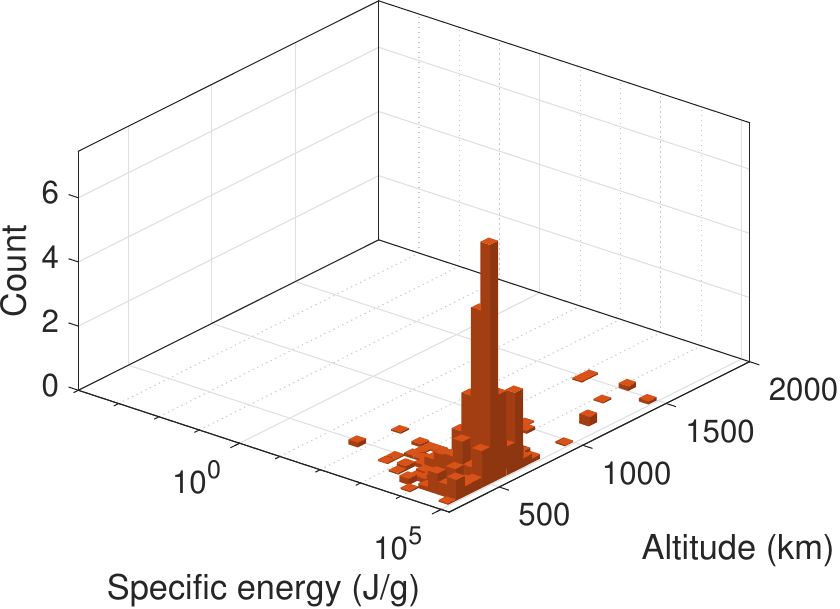}}  
	\caption{Specific energy of collisions per altitude for a subset of collision population for $L_C = 0.2$ cm}
	\label{fig:mega700-fi5-02mm}
\end{figure}

The NASA SBM categorizes a collision as catastrophic or non-catastrophic with a simple specific energy threshold $\Tilde{E}_p^*=40$ J/g threshold \cite{nasaSTMevolve4}.  For context, a 50 g golf ball hit at 42 m/s will yield 44 Joules, and such a golf ball hitting another stationary golf ball would produce a specific energy of around 0.9 J/g.  
For a typical collision geometry with  $v_{imp} \approx 11$ km/s, the specific energy threshold of 40 J/g is reached when the ratio of the two masses is greater than $\approx$ 1500:1.  Therefore, for a catastrophic collision to occur given an LEO payload mass, the colliding object must have some minimum mass.  This means that the LNTs generally only produce noncatastrophic collisions, and the cascading effect of on-orbit collisions is reduced.  This effect is more clearly seen when the histogram of collision energy is shown for a range of $L_C$, as shown in Fig. \ref{fig:mega700fi5sStackHist}. Each scenario is an median value of 20 MC runs over a 100-year simulation, with launches of megaconstellations below 700 km.  The red line denotes the $\Tilde{E}_p^*$ threshold.  In this figure, the collision types are divided into three groups: object vs. object, object vs. debris, and debris vs. debris.  Object here is defined as all intact objects, which includes derelict and rocket body objects.  Active payloads are not represented in these collisions as $\alpha_a = 0$ for these simulations. Debris is defined as all other objects.  The red line denotes the 40 J/g threshold for catastrophic collision, as defined by the NASA SBM.  As shown before, smaller $L_C$ simulations show more collisions, and the collision rate for noncatastrophic collisions grows much more quickly. When simulations only consider objects greater than 10 cm, the collision rate between noncatastrophic and catastrophic collisions may seem comparable, but a higher fidelity simulation shows that the number of noncatastrophic collisions may be much higher and is sensitive to the $L_C$ considered in the simulation.  The histogram of specific energy of the collisions in different altitude shells for three $L_C$ values are shown in Figs. \ref{fig:mega700-fi5-10cm}-\ref{fig:mega700-fi5-02mm}, where the red region denotes collisions that are catastrophic.

\begin{figure}
    \centering
    \includegraphics[width=.5\linewidth]{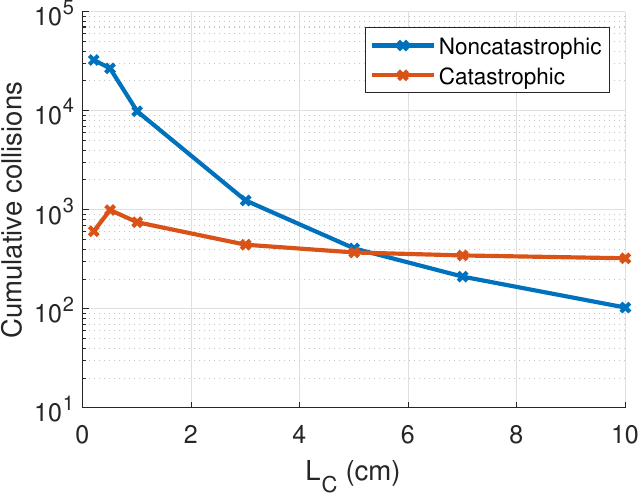}
    \caption{Cumulative number of collision for the 700 km megaconstellation launch case after 100 years for $L_C$ values from 0.2 cm to 10 cm}
    \label{fig:mega700-LCvsNcol}
\end{figure}

The distinct difference in the peaks of the object versus debris and debris vs debris collision energy is explained by the definition of the specific energy, which is soley by the impact velocity and the ratio of masses between the two parent objects.  The specific energy of a collision between two identical masses -- whether its a 10 g debris or a 1000 kg payload -- yield the same specific energy.  The peaks of the histograms therefore correspond to the ratio of the collision masses when that pair of objects collide, assuming similar impact velocity.  Objects tend to be much more massive compared to a debris, and therefore the specific energy is reduced for the object-on-object collisions compared to object-on-debris collisions.  This is also seen by the peak of the debris-on-debris collisions reducing in its specific energy as $L_C$ is reduced -- there is far more collisions between two unequal masses within the debris class as smaller debris are simulated in an MC environment.  $L_C$ = 10 cm case on average had 316.7 catastrophic collisions out of 409.9 total collisions, $L_C$ = 1 cm had 480.4 of 6760.0, and $L_C$ = 0.2 cm had 405.1 of 23773.1 collisions that were catastrophic. This relationship is summarized in Fig. \ref{fig:mega700-LCvsNcol}.  

Note that while it is true that a catastrophic collision would produce more objects compared to a non-catastrophic collision between the same pair of objects, not all of the catastrophic collisions shown in these histograms would necessarily produce more debris than the non-catastrophic collision.  The number of debris is ultimately determined by the masses of the parent objects, and a catastrophic collision between two objects would produce more debris than one between two debris.  Consequently, it is possible that a noncatastrophic collision between an object and debris yields more debris than a catastrophic debris-on-debris collision.

\subsection{The Effect of Improved PMD Rates}

The effect of improved PMD is explored for the various $L_C$ from in the previous section.  The PMD failure rate has been halved, as indicated in Table \ref{tab:improvedPMD}.  

\begin{table}[htp!]
\centering
\caption{Historical and Improved PMD Rates}
\begin{tabular}
  {>{\raggedleft\arraybackslash}p{4cm}%
   >{\centering\arraybackslash}p{4cm}%
   >{\centering\arraybackslash}p{4cm}%
  }
\toprule
  & Historical PMD \cite{ESAspaceEnvReport2023}  &  Improved PMD \\  
 \midrule
Active (non-constellation) & 40\% & 70\% \\ 
Active (constellation) & 90\% & 95\% \\
Rocket Body & 55\% & 77.5\% \\  \bottomrule
\end{tabular} 
\label{tab:improvedPMD}
\end{table}

\begin{figure}
    \centering
    \includegraphics[width=0.5\textwidth]{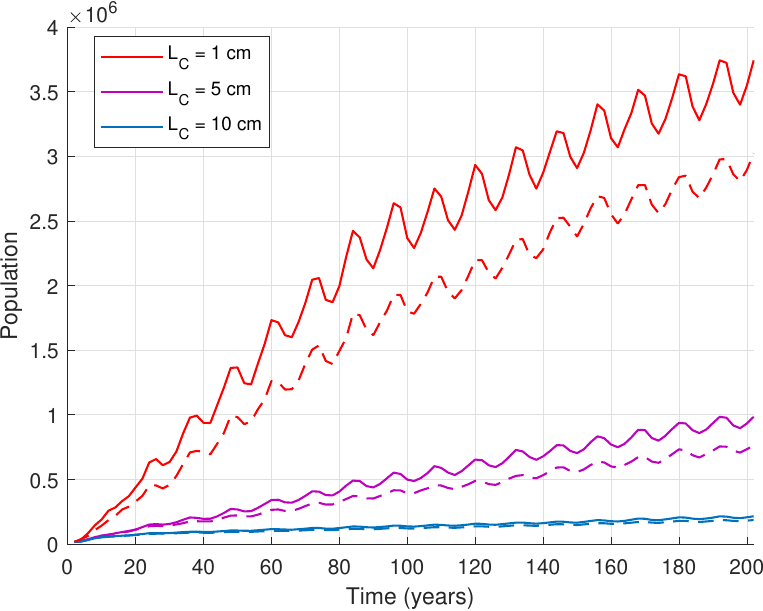}
    \caption{Total population count with megaconstellations launches limited to $<700$ km over a 200-year span for
various minimum LNT sizes and with increased PMD efficacy. Dotted lines represent the high PMD cases. The altitude bins are 50 km.}
    \label{fig:mega700LNT_pops_PMD}
\end{figure}

\begin{figure}[!ht]
    \centering
	\subcaptionbox{Catastrophic collisions}[.45\textwidth]{\includegraphics[width=0.9\linewidth]{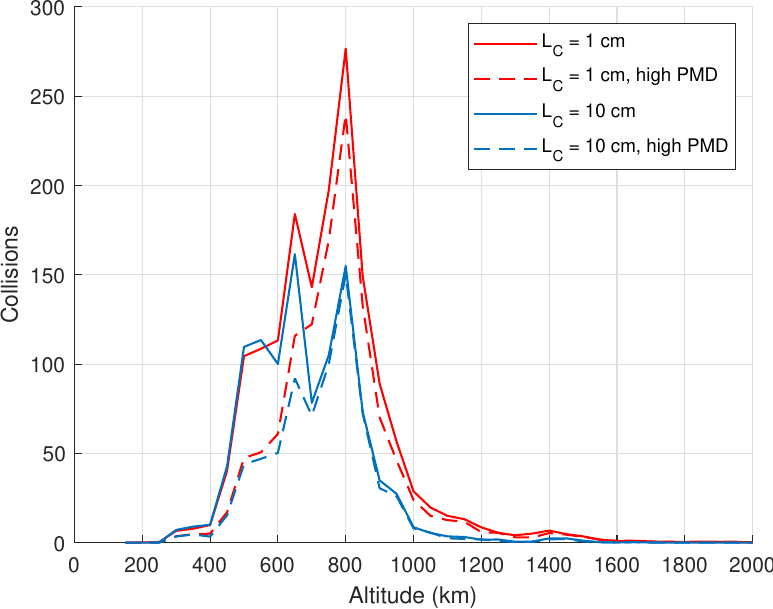}}
	\subcaptionbox{All collisions}[.45\textwidth]{\includegraphics[width=0.9\linewidth]{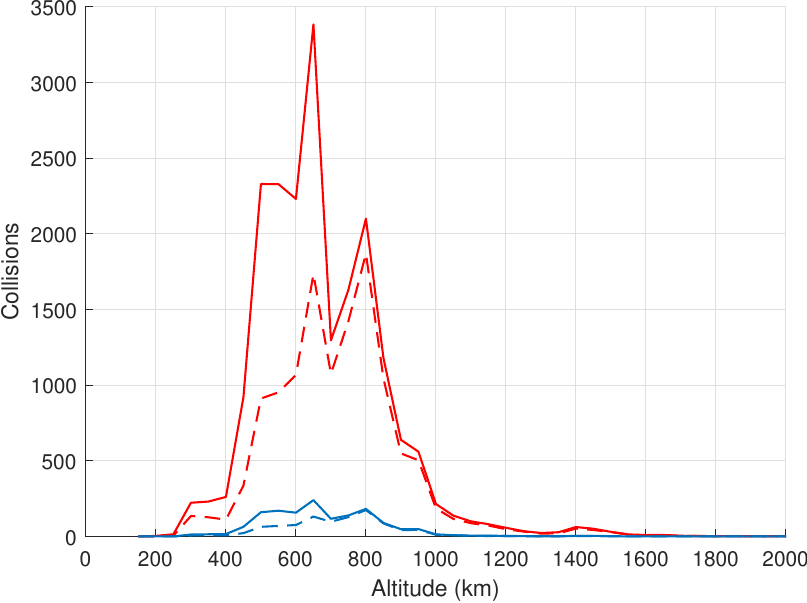}}  
	\caption{Cumulative number of collisions with megaconstellations launches limited to $<700$ km over a 200-year span for  various minimum LNT sizes and with improved PMD}
	\label{fig:mega700LNT_col_PMD}
\end{figure}

Note that the vast majority of launches in this scenario comprise constellation payloads.  The reduction in the total orbital population due to this is shown in Fig. \ref{fig:mega700LNT_pops_PMD}.  Because the LNT population depends heavily on the number of collisions, there is a greater effect in controlling the LNT population with the reduction of derelict objects.  The reduction in the collision rate is also seen in Fig. \ref{fig:mega700LNT_col_PMD}.  With the amount of derelict produced per year effectively halved, the number of catastrophic collisions for altitudes with launches ($<700$ km) is halved.  This is seen for both the scenario with $L_C$ = 1 cm and 10 cm.  The effect on non-catastrophic collision is even greater, as the opportunity for debris-on-debris collision is also reduced.  However, note that for any type of collision, improved PMD can only effectively reduce the collisions in the altitude regions where the launches occur.  The altitude regions without launches do see some reduction due to the reduction in eccentric debris created from collisions.  These findings are in line with the literature, where larger and higher objects seem to be the most polluting \cite{mcknight2021, MITRI}.

\begin{figure}[!ht]
    \centering
	\subcaptionbox{Historic PMD}[.45\textwidth]{\includegraphics[width=0.9\linewidth]{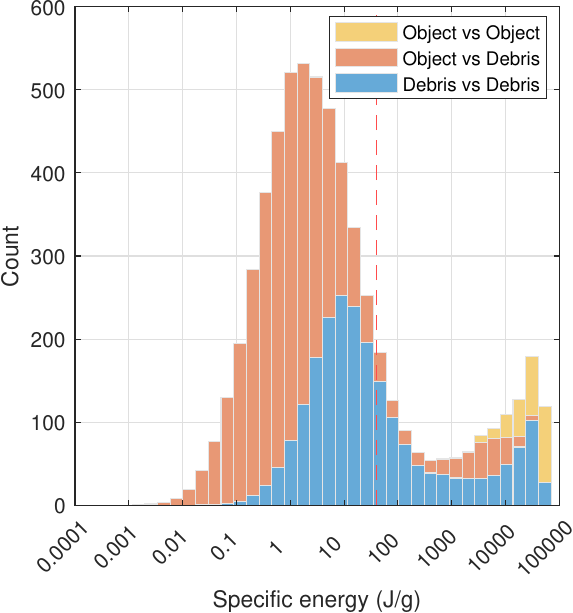}}
	\subcaptionbox{High PMD}[.45\textwidth]{\includegraphics[width=0.9\linewidth]{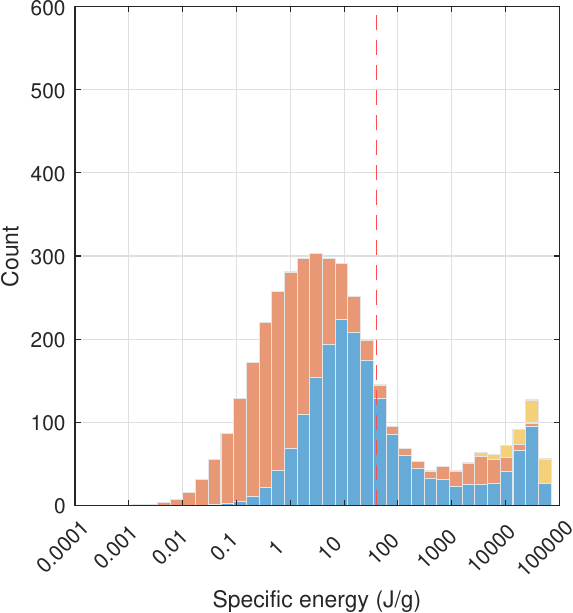}}  
	\caption{Effect of improved PMD on specific energy of collisions for limited to $<700$ km megaconstellation launch scenario for minimum characteristic length $L_C$ = 3 cm}
	\label{fig:mega700-fi5-03cmPMDplus}
\end{figure}

Figure \ref{fig:mega700-fi5-03cmPMDplus} compares the cumulative collision occurance binned into specific energy between the historic PMD case and the high PMD case.  Most of the reduction in collisions is seen in the non-catastrophic collisions due to the higher rate of successful PMD, though a reduction in the high energy object-on-object collisions is also shown.  The reduction in the secondary collision between smaller objects is shown to be a by-product of higher PMD rates, which can arise from rigorous policy efforts and technological improvements.  

\clearpage

\section{Conclusions}
This work explores the effect of including lethal non-trackable objects in the evolution of the LEO environment.  Most orbital objects today are untracked or untrackable due to the limitations of the sensors as measured by orbital in-situ sensors, yet these have not been modeled using high-fidelity evolutionary models due to the large amount of computational resources required to propagate millions of objects.  MOCAT-MC has been modified to efficiently simulate the evolution of the orbital population, which includes LNTs and many future constellations.  LNTs require special consideration as detecting such objects is challenging, and thus the probability of a successful collision avoidance maneuver is lower.  This LNT model was validated against the ADEPT model and presented in this paper, and good initial population reduction and collision statistics are shown down to 2 cm objects, though the differences between the breakup models need further investigation.  Using MOCAT-MC, LNTs as small as 2 mm are modeled for the first time in literature and its effect on the increased collision rates is quantified.  Possible future improvements to PMD rates with the inclusion of LNTs is presented and discussed.  Increased PMD shows a large effect in curbing the non-catastrophic collisions in simulations with LNTs and reducing the general population growth.  

It is shown through several scenarios that while the population of LNTs follows the power-law distribution, the number of catastrophic collisions does not grow linearly with the LNT population, as most of these objects do not collide with larger rocket bodies, derelict and payload objects with sufficient specific energy.  However, the impact on payload survivability is pronounced, and the creation of derelict objects can cause secondary catastrophic collisions.  Similar to debris at 10 cm ranges, the LNT population also shows a steep drop in population density below 500 km altitudes due to atmospheric drag. 

\section*{Acknowledgments}
Research was sponsored by the United States Air Force Research Laboratory and the Department of the Air Force Artificial Intelligence Accelerator and was accomplished under Cooperative Agreement Number FA8750-19-2-1000. The views and conclusions contained in this document are those of the authors and should not be interpreted as representing the official policies, either expressed or implied, of the Department of the Air Force or the U.S. Government. The U.S. Government is authorized to reproduce and distribute reprints for Government purposes notwithstanding any copyright notation herein.
This material is based upon work supported by the National Science Foundation Graduate Research Fellowship Program under Grant No. 2141064. Any opinions, findings, and conclusions or recommendations expressed in this material are those of the author(s) and do not necessarily reflect the views of the National Science Foundation.
Daniel Jang is supported by the MIT Lincoln Laboratory Scholarship Program Fellowship.

The authors thank Miles Lifson and The Aerospace Corporation for providing the ADEPT dataset used for validation.  

The authors also thank the MIT SuperCloud and Lincoln Laboratory Supercomputing Center for providing HPC, database, and consultation resources that have contributed to the research results reported in this paper.  

\clearpage
\bibliography{referencesThesis}
\end{document}